\documentclass{ws-mpla}
\def\be{\begin{equation}}
\def\ee{\end{equation}}
\def\bea{\begin{eqnarray}}
\def\eea{\end{eqnarray}}

\usepackage{graphicx}
\usepackage{dcolumn}
\usepackage{bm}

\begin{document}

\markboth{Authors' Names}
{Instructions for Typing Manuscripts (Paper's Title)}

\catchline{}{}{}{}{}

\title{Extended Schouten classification for non-Riemannian geometries
}

\author{\footnotesize SABRINA CASANOVA}

\address{Max Planck Institut fuer Kernphysik,\\ Saupfercheckweg 1, 69117 Heidelberg, Germany\\ICRA, Piazzale Aldo Moro 5, 00185, Rome, Italy\\
Sabrina.Casanova@mpi-hd.mpg.de}

\author{ORCHIDEA MARIA LECIAN}

\address{Universita' degli Studi di Roma La Sapienza and ICRA,\\ Piazzale Aldo Moro 5, 00185, Rome, Italy
}

\author{GIOVANNI MONTANI}

\address{Universita' degli Studi di Roma La Sapienza and ICRA,\\ Piazzale Aldo Moro 5, 00185, Rome, Italy\\ENEA C.R. Frascati,\\ Via Enrico Fermi 45, 00044 Frascati, Italy\\ICRANET,\\ P.le della Repubblica, 65122, Pescara, Italy
}

\author{REMO RUFFINI}

\address{Universita' degli Studi di Roma La Sapienza and ICRA,\\ Piazzale Aldo Moro 5, 00185, Rome, Italy\\ICRANET,\\ P.le della Repubblica, 65122, Pescara, Italy
}

\author{ROUSTAM ZALALETDINOV}

\address{ICRANET,\\ P.le della Repubblica, 65122, Pescara, Italy
}

\maketitle

\pub{Received (Day Month Year)}{Revised (Day Month Year)}

\begin{abstract}
A generalized connection, including Christoffel coefficients, torsion, non-metricity tensor and metric-asymmetricity object, is analyzed according to the Schouten classification. The inverse structure matrix is found in the linearized regime, autoparallel trajectories are defined and the contribution of the components of the connection are clarified at first-order approximation. 

\end{abstract}

\ccode{PACS Nos.: 12.10.-g}

\section{Introduction}

  This paper is aimed at proposing an approach to a classification of affine-connection geometries with an asymmetric metric tensor, according to the Schouten scheme, starting from the definition of all objects and finding the compatibility conditions between them. The overwhelming majority of the approaches to geometries with an asymmetric metric (see \cite{Scho:1954} for a review of classical results) are physically motivated by starting from a Lagrangian to derive both field equations and the definition of the connection in terms of an asymmetric metric, by means of a Palatini variational principle. However, such an approach is restricted from the very beginning by a fixed Lagrangian, able to provide only a fixed class of geometries. Alternatively, a classification of such geometries in analogy with that of Schouten \cite{Scho-Stru:1935} for the class of affine-connection geometries with a symmetric metric would provide a proper understanding of the structure and variety of possible geometries with an asymmetric metric.\\
In a generalized Schouten classification, the connection consists of four components, i.e., Christoffel coefficients, the metric-asymmetricity object, generalized contortion tensor, and non-metricity tensor. The role of these components can be analyzed according to the different aspects of physics that are to be investigated. For a comparison between the macroscopic and the microscopic approaches in the case of torsion, see \cite{nak}.\\
In \cite{drago}, two mechanical (macroscopic) examples are discussed, the case of a homogeneous disc rolling without sliding on a horizontal plane, and that of a homogeneous ball rolling without sliding on a sphere. Here, the non-holomic connection introduced by Schouten is analyzed in the context of Wagner's proposal for the curvature tensor.\\
On the other hand, two main research lines can be outlined from a quantum point of view: that of Unified-Field theories, and that of Geometry-plus-Matter theories. In general, these two approaches make different predictions.\\
In the first case, (see \cite{eddi} and \cite{einst} for the earliest attempts, and \cite{rev} for a recent review), all fields are geometrical.  In \cite{rob}, torsion is assumed vanishing, because it is linked with spin \cite{hehvdhker76}, and matter is related with the non-symmetric part of the metric, while dilaton with the non-metricity object.\\
In the second case, the coupling of matter fields with geometry has to be evaluated, and the features of the connection components have to be interpreted. In \cite{popla}, the connection is established to be antisymmetric in the first two indices only for a metric-compatible affine connection, and covariant differentiation for spinor fields is investigated. The correlation between torsion and Yang-Mills fields is widely explored in \cite{miel}: the mapping between these two kinds of fields is discussed in terms of of differential geometry. As a result, Riemann-Cartan geometries are shown to reproduce Yang-Mills equations, while Yang-Mills connections are illustrated to induce richer structures.\\
After reviewing the Schouten classification, we will find the inverse of the ''structure matrix'', which links the generalized connection with all the metric objects, in the linear approximation. The expression of autoparallel trajectories will then be evaluated in the first-order approximation. Brief concluding remarks follow.

\section{Schouten classification}

  According to the Schouten classification \cite{Scho-Stru:1935}, a
non-Riemannian geometry of general setting, with a symmetric metric tensor $
g_{\mu \nu }$ and two different affine connections, $\Pi _{\nu \sigma }^{\mu
}$ for the parallel transportation of covectors $a_{\mu }$ and $\Theta _{\nu
\sigma }^{\mu }$ for the parallel transportation of vectors $v^{\mu }$, is
characterized by the presence of three tensors of rank (1,2) which are
responsible for the non-Riemannian character:
\begin{itemlist}
 \item a difference tensor
between the two connections $\label{difference}S_{\nu \sigma }^{\mu }=\Pi
_{\nu \sigma }^{\mu }-\Theta _{\nu \sigma }^{\mu } $;
 \item a torsion
tensor $\label{torsion}T_{\nu \sigma }^{\mu }=2\Pi _{[\nu \sigma ]}^{\mu }$
(square brackets denote antisymmetrisation)
  \item a
non-metricity object\footnote{where $|$ denotes covariant derivation with respect to the connection $\Theta$} $\label{non-metricity}N_{\nu \sigma }^{\mu }=g^{\mu
\varepsilon }g_{\nu \sigma \mid \varepsilon }$ due to the incompatibility of
metric and connection in general.
\end{itemlist}
From the definition of the non-metricity object, one gets
the connection ${\Theta }_{\nu \sigma
}^{\mu }$
\begin{equation}
{\Theta }_{\nu \sigma
}^{\mu }={\Gamma }_{\nu \sigma }^{\mu }+A_{\nu \sigma }^{\mu },
\end{equation}
where
\begin{equation} {
\Gamma }_{\nu \sigma }^{\mu }=\frac{1}{2}g^{\mu \varepsilon }(g_{\nu \varepsilon ,\sigma
}+g_{\sigma \varepsilon ,\nu }-g_{\nu \sigma ,\varepsilon })
\end{equation}
is the metric connection, and $A_{\nu
\sigma }^{\mu }$ is the so-called affine-deformation tensor, defined as 
\begin{equation}
A_{\nu \sigma }^{\mu }=-S_{\nu \sigma }^{\mu }+C_{\nu \sigma }^{\mu }-D_{\nu
\sigma }^{\mu },
\end{equation}
\begin{subequations}\label{dynamical equations}
\begin{align}\label{Einstein total equations}
&C_{\nu \sigma }^{\mu }=\frac{1}{
2}\left[ g^{\rho \mu }(T_{\sigma \rho }^{\varepsilon }g_{\varepsilon \nu
}+T_{\nu \rho }^{\varepsilon }g_{\varepsilon \sigma })+T_{\nu \sigma
}^{\varepsilon }g_{\varepsilon }^{\mu }\right]
\\
&D_{\nu \sigma }^{\mu }=\frac{1}{2}g^{\mu \varepsilon }(N_{\nu \varepsilon
,\sigma }+N_{\sigma \varepsilon ,\nu }-N_{\nu \sigma ,\varepsilon })
\end{align}
\end{subequations}
being the contortion tensor and the non-metricity tensor, respectively.\\
The case of Riemannian geometry is the totally degenerate case,
$\label{Riemannian}S_{\nu \sigma }^{\mu }=T_{\nu
\sigma }^{\mu }=N_{\nu \sigma }^{\mu }=0$.\\

\subsection{Curvature tensor}

  The most important result of the Schouten classification is that the curvature
tensor $R^{\alpha }{}_{\beta \rho \sigma }$ for the
connection $\Theta _{\nu \sigma }^{\mu }$ contains the
Riemannian curvature $M^{\alpha }{}_{\beta \rho \sigma }$ of the metric
connection ${\Gamma }_{\nu \sigma }^{\mu }$, decoupled from non-Riemannian
contributions,
\begin{equation}
R^{\alpha }{}_{\beta \rho \sigma }=M^{\alpha }{}_{\beta \rho \sigma
}+2A^{\alpha }{}_{\beta [\sigma ;\rho ]}+2A^{\alpha }{}_{\varepsilon [\rho
}A^{\varepsilon }{}_{\underline{\beta }\sigma ]}  \label{structure}
\end{equation}
 where $;$ denotes the
covariant derivative with respect to ${\Gamma }_{\nu \sigma }^{\mu }$ (underlined indices are not affected by antisymmetrisation). The
decoupling of the Riemannian part from the non-Riemannian one is important
for the analysis of extra structures on non-Riemannian spaces,
especially in cases when such structures are known not to exist on
Riemannian manifolds, so that they may be expected to be compatible with
objects of non-Riemannian character. From the physical point of
view, it means that a field theory based on such a non-Riemannian geometry
always contains Riemannian gravity (general relativity with an appropriate
Lagrangian) and extra fields as non-Riemannian (non-gravitational) effects: a non-Riemannian geometry always contains
the Riemannian part, as represented in general
relativity, with its Lagrangian, and extra fields
due to the non-Riemannian structures.

\section{Generalized Schouten classification of non-Riemannian geometries with an asymmetric metric tensor}

  Let us now consider now an affine-connection geometry, with connections $\Pi _{\nu
\sigma }^{\mu }$ and $\Theta _{\nu \sigma }^{\mu }$, as described above, and
with an asymmetric metric tensor $g_{\mu \nu }$, $g_{\mu \nu }\neq g_{\nu
\mu }$. A generalized Schouten classification can be constructed for this case.
An asymmetric metric tensor $g_{\mu \nu }$ can always be split into its
symmetric part $s_{\mu \nu }$ and antisymmetric part $a_{\mu \nu }$,
\begin{equation}\label{qwert}
g_{\mu \nu }=s_{\mu \nu }+a_{\mu \nu }.
\end{equation}
The metric tensor $s_{\mu \nu }$ is
assumed to lower and raise tensor indices together with its inverse $
s^{\mu \nu }$. Similarly to the case of the symmetric metric, the analysis of the
incompatibility between metric and connection\footnote{from now on, $\mid$ will denote covariant derivation with respect
to the metric part of the connection.} $g_{\mu \nu \mid \rho }=N_{\mu
\nu \rho }$ brings about the following expression for the connection $\Pi
_{\kappa \lambda }^{\theta },$
\begin{equation}
\Pi _{\kappa \lambda }^{\theta }(\delta _{\theta }^{\sigma }\delta _{\nu
}^{\kappa }\delta _{\rho }^{\lambda }+g^{\sigma \lambda }\delta _{\rho
}^{\kappa }a_{\theta \nu }+g^{\sigma \kappa }\delta _{\nu }^{\lambda
}a_{\rho \theta })=\Gamma _{\nu \rho }^{\sigma }+\Delta _{\nu \rho }^{\sigma
}+{\bf C}_{\nu \rho }^{\sigma }-{\bf D}_{\nu \rho }^{\sigma }\,,
\label{eqndet}
\end{equation}
where $\Gamma _{\nu \rho }^{\sigma }$ is 
the usual Christoffel connection, while
\begin{subequations}\label{lkjh}
\begin{align}
&\Delta _{\nu \rho
}^{\sigma }=\frac{1}{2}s^{\sigma \mu }(a_{\mu \nu ,\rho }+a_{\rho \mu ,\nu
}-a_{\nu \rho ,\mu })
\\
&C_{\nu \rho
}^{\sigma }=\frac{1}{2}\left[ s^{\sigma \mu }(T_{\nu \mu }^{\varepsilon
}g_{\varepsilon \rho }+T_{\rho \mu }^{\varepsilon }g_{\varepsilon \nu
})+T_{\nu \rho }^{\varepsilon }g_{\varepsilon }^{.\sigma }\right] 
\\
&D_{\nu \rho }^{\sigma }=\frac{1}{2}s^{\sigma \mu
}(N_{\mu \nu \rho }+N_{\rho \mu \nu }-N_{\nu \rho \mu })
\end{align}
\end{subequations}
are the metric-asymmetricity object, the generalized contortion tensor, and the non-metricity tensor, respectively, specified for the metric (\ref{qwert}).
The determinant
of the ''hypercubic'' {\em structure matrix} $J_{\theta \nu \rho }^{\sigma
\kappa \lambda }$, 
\begin{equation}
J_{\theta \nu \rho }^{\sigma
\kappa \lambda }=\delta _{\theta }^{\sigma }\delta _{\nu }^{\kappa }\delta
_{\rho }^{\lambda }+g^{\sigma \lambda }\delta _{\rho }^{\kappa }a_{\theta
\nu }+g^{\sigma \kappa }\delta _{\nu }^{\lambda }a_{\rho \theta },
\end{equation}
is related to the existence of solutions of the system of {\em inhomogeneous}
{\em linear} {\em algebraic} equations (\ref{eqndet}) for the unknowns $\Pi
_{\beta \gamma }^{\alpha }$ similar to the case of usual quadratic matrices.

\subsection{Generalized curvature tensor}

  When the determinant is different from zero, the system has a non-trivial
solution, which can be expressed through the inverse structure matrix $
\widetilde{J}_{\sigma \beta \gamma }^{\alpha \nu \rho }=
{(J^{-1})}_{\sigma \beta \gamma }^{ \alpha \nu \rho }$,
$\widetilde{J}_{\sigma \beta \gamma
}^{\alpha \nu \rho }J_{\mu \varepsilon \lambda }^{\sigma \beta \gamma
}=\delta _{\mu }^{\alpha }\delta _{\varepsilon }^{\nu }\delta _{\lambda
}^{\rho }$, as
\begin{equation}
\Pi _{\beta \gamma }^{\alpha }=(\Gamma _{\nu \rho }^{\sigma }+\Delta _{\nu
\rho }^{\sigma }+{\bf C}_{\nu \rho }^{\sigma }-{\bf D}_{\nu \rho }^{\sigma })
\widetilde{J}_{\sigma \beta \gamma }^{\alpha \nu \rho }.  \label{eqndetbis}
\end{equation}
This way, the curvature tensor for $\Pi _{\beta \gamma }^{\alpha }$ can be written
in the form
\begin{equation}
R^{\alpha }{}_{\beta \rho \sigma }=\widetilde{M}^{\alpha }{}_{\beta \rho
\sigma }+2\widetilde{A}^{\alpha }{}_{\beta [\sigma \Vert \rho ]}+2\widetilde{A}
^{\alpha }{}_{\varepsilon [\rho }\widetilde{A}^{\varepsilon }{}_{\underline{
\beta }\sigma ]},  \label{structure2}
\end{equation}
where ${\widetilde{M}}^{\alpha }{}_{\beta \rho \sigma }$ is the curvature tensor
for an affine connection $\widetilde{\Gamma }_{\beta \gamma }^{\alpha }
=(\Gamma _{\nu \rho }^{\sigma }+\Delta _{\nu \rho }^{\sigma })\widetilde{J}
_{\sigma \beta \gamma }^{\alpha \nu \rho }$, $\widetilde{A}_{\beta \gamma
}^{\alpha }=(C_{\nu \rho }^{\sigma }-D_{\nu \rho }^{\sigma })\widetilde{J}
_{\sigma \beta \gamma }^{\alpha \nu \rho }$ is a generalized affine
deformation tensor and $\Vert $ is the covariant derivative with respect to $
\widetilde{\Gamma }_{\beta \gamma }^{\alpha }$. Eq. (\ref
{structure2}) is a generalisation of (\ref{structure}) for the case of an asymmetric
metric. The extraction of the Riemannian curvature tensor $M^{\alpha }{}_{\beta \rho
\sigma }$ from $\widetilde{M}
^{\alpha }{}_{\beta \rho \sigma }$ gives a direct analogue of (\ref{structure}
)
\begin{equation}
R^{\alpha }{}_{\beta \rho \sigma }=M^{\varepsilon }{}_{\nu \rho \lambda }
\widetilde{J}_{\varepsilon \beta \sigma }^{\alpha \nu \lambda }+\Sigma
^{\alpha }{}_{\beta \rho \sigma }(\widetilde{A}^{\alpha }{}_{\beta \sigma
},\Delta ^{\alpha }{}_{\beta \sigma },\widetilde{J}_{\sigma \beta \gamma
}^{\alpha \nu \rho }),  \label{structure3}
\end{equation}
where $\Sigma ^{\alpha }{}_{\beta \rho \sigma }$ is a {\em tensor}
constructed from generalised affine deformation tensor, metric asymmetricity
object and the inverse structure matrix and their derivatives.

\subsection{Linear approximation}

  The determinant of ${J_{\theta \nu
\rho }^{\sigma \kappa \lambda }}$ has been calculated in a perturbation
expansion in terms of small asymmetric metric, $\mid
a_{\mu \nu }\mid \ll \mid s_{\mu \nu }\mid $. Then, in linear approximation,
the matrix
\begin{equation}\label{lin} 
J{{_{\theta \nu \rho }^{\sigma \kappa \lambda }}}={\delta }
_{\theta }^{\sigma }{\delta }_{\nu }^{\kappa }{\delta }_{\rho }^{\lambda
}+s^{\sigma \lambda }{\delta }_{\rho }^{\kappa }a_{\theta \nu }+s^{\sigma
\kappa }{\delta }_{\nu }^{\lambda }a_{\rho \theta }
\end{equation} 
has its inverse as 
\begin{equation}\label{invlin}
\widetilde{J}_{\sigma \beta \gamma }^{\alpha \nu \rho }={\delta }_{\sigma
}^{\alpha }{\delta }_{\beta }^{\nu }{\delta }_{\gamma }^{\rho }-s^{\alpha
\nu }{\delta }_{\beta }^{\rho }a_{\gamma \sigma }-s^{\alpha \rho }{\delta }
_{\gamma }^{\nu }a_{\sigma \beta }.
\end{equation}
Eq.s (\ref{eqndet})-(\ref
{structure3}) are the main relations describing the structure of affine-connection geometries with an asymmetric metric.

\section{Autoparallel trajectories}

  So far, it is possible to analyze the expression of autoparallel trajectories. Be $u^{\mu}=dx^{\mu}/d\lambda$ the tangent vector to the curve $x^{\mu}=x^{\mu}(\lambda)$. Because of the solution (\ref{invlin}), the autoparallel equation
\begin{equation}\label{autop}
\frac{du^\alpha}{d\lambda}+\Pi^\alpha_{\beta\gamma}u^\beta u^\gamma=0
\end{equation}
simplifies as
\begin{equation}
\frac{du^\alpha}{d\lambda}+\left(\Gamma^{\sigma}_{\nu\rho}+\Delta^{\sigma}_{\nu\rho}+C^{\sigma}_{\nu\rho}-D^{\sigma}_{\nu\rho}\right)\tilde{J}^{\alpha\nu\rho}_{\sigma\beta\gamma}u^\beta u^\gamma=0.
\end{equation}
Because of the symmetries of (\ref{invlin}), (\ref{autop}) rewrites
\begin{align}
&\frac{du^\alpha}{d\lambda}+\Gamma^{\alpha}_{\nu\rho}u^\nu u^\rho+
\Delta^{\sigma}_{\nu\rho}
\left(-s^{\alpha\nu}a_{\gamma\sigma}u^{\rho}u^{\gamma}+s^{\alpha\rho}a_{\beta\sigma}u^{\nu}u^{\beta}\right)+\nonumber\\
&+\frac{1}{2}s^{\mu\nu}\left(T^{\epsilon}_{\nu\mu}g_{\epsilon\rho}
+T^{\epsilon}_{\rho\mu}g_{\epsilon\nu}\right)\delta^{\alpha}_{\sigma}u^{\rho}u^{\sigma}+
\frac{1}{2}T^{\epsilon}_{\nu\rho}g_{\epsilon}^{\ \ \sigma}\left(-s^{\alpha\nu}a_{\gamma\sigma}u^{\rho}u^{\gamma}+s^{\alpha\rho}a_{\beta\sigma}u^{\nu}u^{\beta}\right)+\nonumber\\
&-\frac{1}{2}s^{\sigma\mu}\left(N_{\mu\nu\rho}+N_{\rho\mu\nu}-N_{\nu\rho\mu}\right)\left(\delta^{\alpha}_{\sigma}u^{\nu}u^{\rho}-s^{\alpha\nu}a_{\gamma\rho}u^{\gamma}u^{\rho}-s^{\alpha\rho}a_{\sigma\beta}u^{\beta}u^{\nu}\right)=0.
\end{align}
If we expand the previous expression and keep only the terms linear in $a$, we obtain
\begin{align}
&\frac{du^\alpha}{d\lambda}+\Gamma^{\alpha}_{\nu\rho}u^\nu u^\rho+\frac{1}{2}s^{\mu\nu}\left(T^{\epsilon}_{\nu\mu}s_{\epsilon\rho}
+T^{\epsilon}_{\rho\mu}s_{\epsilon\nu}\right)\delta^{\alpha}_{\sigma}u^{\rho}u^{\sigma}+
\nonumber\\
&+\frac{1}{2}s^{\mu\nu}\left(T^{\epsilon}_{\nu\mu}a_{\epsilon\rho}
+T^{\epsilon}_{\rho\mu}a_{\epsilon\nu}\right)\delta^{\alpha}_{\sigma}u^{\rho}u^{\sigma}+
\frac{1}{2}T^{\epsilon}_{\nu\rho}s_{\epsilon}^{\ \ \sigma}\left(-s^{\alpha\nu}a_{\gamma\sigma}u^{\rho}u^{\gamma}+s^{\alpha\rho}a_{\beta\sigma}u^{\nu}u^{\beta}\right)+\nonumber\\
&-\frac{1}{2}s^{\sigma\mu}\left(N_{\mu\nu\rho}+N_{\rho\mu\nu}-N_{\nu\rho\mu}\right)\left(\delta^{\alpha}_{\sigma}u^{\nu}u^{\rho}-s^{\alpha\nu}a_{\gamma\rho}u^{\gamma}u^{\rho}-s^{\alpha\rho}a_{\sigma\beta}u^{\beta}u^{\nu}\right)=0.
\end{align}
Assuming that $a\sim T\sim N$, we find, at first order,
\begin{equation}
\frac{du^\alpha}{d\lambda}+\Gamma^{\alpha}_{\nu\rho}u^\nu u^\rho+\frac{1}{2}s^{\mu\nu}\left(T^{\epsilon}_{\nu\mu}s_{\epsilon\rho}
+T^{\epsilon}_{\rho\mu}s_{\epsilon\nu}\right)-\frac{1}{2}s^{\sigma\mu}\left(N_{\mu\nu\rho}+N_{\rho\mu\nu}-N_{\nu\rho\mu}\right)\delta^{\alpha}_{\sigma}u^{\rho}u^{\sigma}=0.
\end{equation}
It's interesting to notice that torsion and the non-metricity tensor contribute at this approximation order, while the metric-asymmetricity object provides a negligible contribution.

\section{Conclusions}

  A classification of non-Riemannian geometries with an asymmetric metric tensor has been proposed according to the Schouten classification. By adopting this approach for affine-connection geometries with an asymmetric metric, the structure and 
variety of such geometries can be investigated in a fully-geometrical formalism without adopting any variational 
principle. The definition of autoparallel trajectories at first order has been established: these results can be compared with those of \cite{aprea}, where autoparallel trajectories are derived from a modified Lagrangian, and torsion is shown to be relevant even at zeroth order, playing the same role as the gravitational field.

\section*{Acknowledgments}

  We would like to thank A.A. Kirillov for his valuable advice on some aspects of this subject.


\begin{thebibliography}{0}

\bibitem{Scho:1954} J.A. Schouten, {\it Ricci Calculus} (Springer-Verlag, 1954).

\bibitem{Scho-Stru:1935} J.A. Schouten and D.J. Struik, {\it Einf\"{u}hrung 
in die neueren Methoden der Differentialgeometrie} (Noordhoff, 1935), Vol.~1.

\bibitem{nak} N. Carlevaro, O.M. Lecian and G. Montani, Macroscopic and microscopic paradigms for the torsion field: from the test-particle motion to a Lorentz gauge theory, to appear on {\it Annales de la Fondation Louis de Broglie}, special issue, (invited paper).

\bibitem{drago} V. Dragovic, B. Gajic, {\it Reg. Chaot. Dyn.} {\bf 8}, 105 (2003), math-ph/0304018 and the references therein.

\bibitem{eddi} A. S. Eddington, {\it The Mathematical Theory of Relativity} (Cambridge University Press, 1923).

\bibitem{einst} A. Einstein, {\it Sitzungsber. Preuss. Akad. Wiss.} {\bf 22}, 414 (1925).
 
\bibitem{rev} H. F. M. Goenner, {\it Living Reviews in Relativity} {\bf lrr-2004-2}, (2004).
 
\bibitem{rob} M. D. Roberts, Freiburg im Breisgau, Germany, Physikalisches Institut, Albert-Ludwigs Universit¨at Freiburg, Strings and Unified Field Theory, hep-th/0607118, and the references therein.

\bibitem{hehvdhker76} F. Hehl, P. von der Heyde, G.D. Kerlick, J. Nester, {\it Rev. Mod. Phys.} {\bf 48}, 393 (1976).

\bibitem{popla} N. J. Poplawski, Bloomington, Indiana, USA,Department of Physics, Indiana University, Covariant differentiation of spinors for a general affine connection, arXiv:0710.3982, and the references therein. 
 
\bibitem{miel} E. W. Mielke, Y. N. Obukhov, F. W. Hehl, {\it Phys.Lett. A} { \bf 192}, 153 (1994) ,gr-qc/9407031, and the references therein.

\bibitem{aprea} G. Aprea, G. Montani, R. Ruffini , {\it Int.J.Mod.Phys. D} {\bf 12}, 1875 (2003) ,gr-qc/0401054.

\end{thebibliography}
\end{document}